\documentclass[aip,apl,reprint]{revtex4-1}
\usepackage{graphicx}
\usepackage{bm}

\newcommand\pictc[5]{\begin{figure}[t,b]
                       \centerline{\vspace{-1mm}
                        \includegraphics[width=#1\columnwidth,height=0.7\textheight,keepaspectratio]{#3}}
                       \protect\caption{\protect\label{fig:#4} #5}\vspace{-2mm}
                    \end{figure}            }

\newcommand\pict[4][0.5]{\pictc{#1}{!tb}{#2}{#3}{#4}}

\newcommand\rpict[1]{\ref{fig:#1}}

\begin{document}
\begin{sloppy}

\title{Tunable fishnet metamaterials infiltrated by liquid crystals}

\author{Alexander Minovich, Dragomir N. Neshev, David A. Powell, Ilya V. Shadrivov, and Yuri~S. Kivshar}

\affiliation{Nonlinear Physics Center, Research School of Physics and Engineering,\\Australian National University, Canberra ACT 0200, Australia}

\begin{abstract}
We analyze numerically the optical response and effective macroscopic parameters of fishnet metamaterials infiltrated with a nematic liquid crystal. We show that even a small amount of liquid crystal can provide tuning of the structures due to reorientation of the liquid crystal director. This enables switchable optical metamaterials, where the refractive index can be switched from positive to negative by an external field. This tuning is primarily determined by the shift of the cut-off wavelength of the holes, with only a small influence due to the change in plasmon dispersion.
\end{abstract}


\maketitle

Metamaterials have attracted a great attention due to their unusual electromagnetic properties not available in nature~\cite{Shalaev:2007-41:NatPhot}. One of the important metamaterial designs that is suitable for scaling to optical frequencies is the fishnet metal-dielectric-metal structure~\cite{Zhang:2005-137404:PRL, Dolling:2006-231118:APL, Chettiar:2007-1671:OL, Li:2007-251112:APL, Valentine:2008-376:NAT, Minovich:2010:PRB}.

Due to the strong local field enhancement within the metamaterials, there emerge new opportunities for external control of their properties or changing their operating bandwidth. There are several means to achieve such tunability, including structural~\cite{Lapine:2009-084105:APL}, temperature~\cite{xiao_apl_09, Dicken:2009-18330:OE}, electro-optic~\cite{Zhao:2007-011112:APL, Samson:2009:arXiv}, magneto-optic~\cite{zhang_apl_08}, and nonlinear tuning~\cite{PowShaKiv7, ShaKozWei8}. However, so far there are no works on tuning the optical properties of realistic structures with negative index, such as the fishnet structure.

In this Letter, we numerically analyze the tunability of optical fishnet metamaterials exhibiting negative refractive index. We study a structure with the holes infiltrated by liquid crystals and demonstrate that even a small amount of liquid crystal can allow for substantial tunability of such structures, including reversal of the sign of the refractive index. This is achieved by employing the reorientation of the liquid crystal director using an external field, as well as by implementing the temperature dependence of the liquid crystal refractive index.

Liquid-crystal tunability of metamaterials has been suggested as a path to achieve tunable negative-index metamaterials~\cite{khoo_ol_06}. Subsequent studies have explored this idea in quasi-two-dimensional nanostrip geometry~\cite{wang_apl_07, Werner:2007-3342:OE} and also for  a wire medium~\cite{gorkunov}. Recent experiments demonstrated thermal tunability of optical metamaterials, however the realized structures only experienced a negative magnetic response, and did not show a negative refractive index~\cite{xiao_apl_09}. Therefore, here we present a study on tuning of optical fishnet structures as an important step in achieving tunable negative-index optical metamaterials.

Following our recent experiments~\cite{Minovich:2010:PRB}, we consider a trilayer metal-dielectric-metal structure patterned with a two-dimensional square lattice of rectangular holes through all three layers. The structure is deposited on a glass substrate, as shown in Fig.~\rpict{fig1}. In our simulations, we use realistic material parameters as gold (Au) and zinc-oxide (ZnO) for the metal and the dielectric layers, respectively. We fix the total thickness of the structure, $h=2h_m+h_d=120$\,nm, where $h_m$ is the thickness of the gold and $h_d$ is the thickness of the ZnO layers. We use geometric parameters of the square unit cell of $a=200$\,nm, $b=350$\,nm, and $c=420$\,nm (see Fig.~\rpict{fig1}).

\pict[0.99]{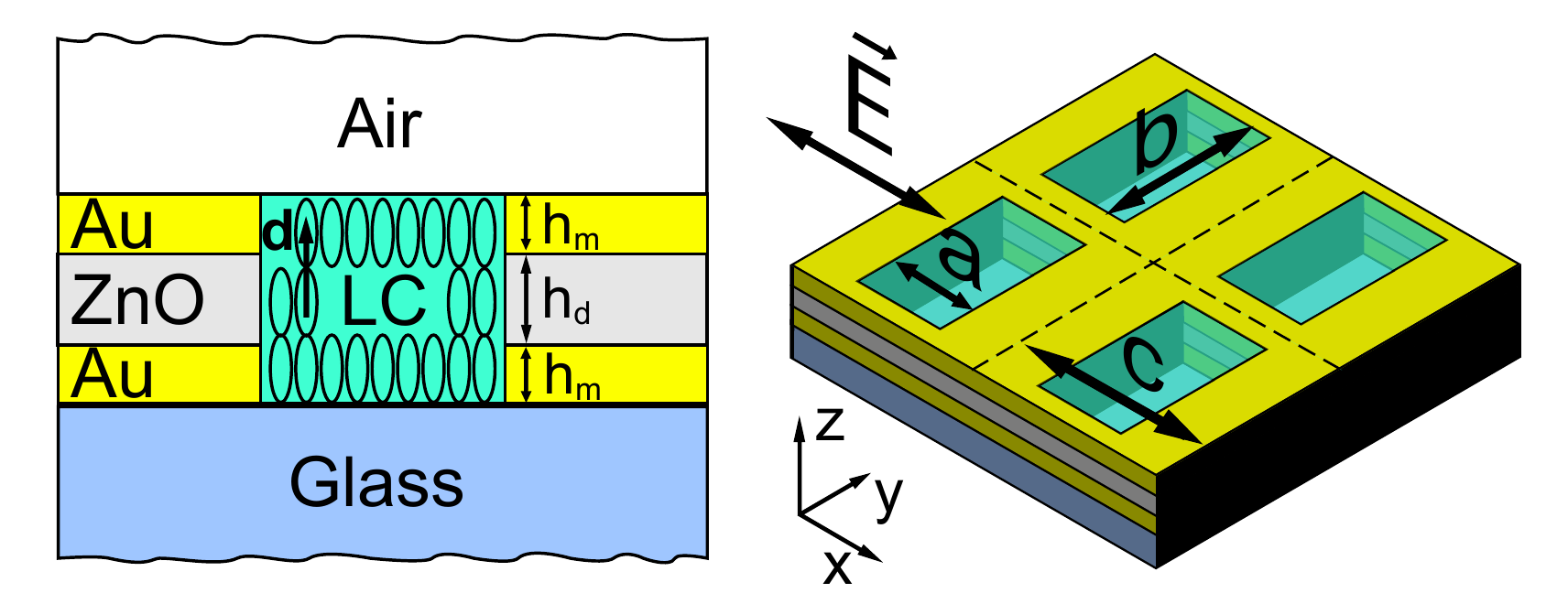}{fig1}{(color online) Trilayer fishnet metamaterial on a glass substrate. Left:  Schematic of the layers. Right: Four unit cells of the structure. $h_m$ is the thickness of metal (Au), $h_d$ is the thickness of dielectric (ZnO). Both substrate and cladding are semi-infinite. Light is polarized along $x$ axis.}

We employ a finite-difference time-domain (FDTD) numerical method performed with commercial software (RSoft). For simplicity, in our simulations we assume that the glass substrate and surrounding air are semi-infinite with refractive indices $n_1=1.48$ and $n_3=1$, respectively. The refractive index of ZnO is taken $n_{\rm ZnO}=1.66$~\cite{Khoshman:2007} and its weak dispersion in the spectral range of $1-2\,\mu$m (maximal deviation from $1.66$ is $1.1\%$) allows for a good approximation by a constant value. For the dielectric permittivity of gold we use six-term Drude-Lorentz expression from the RSoft material library~\cite{Rakic:1998-5271:AO}. A detailed description of the implementation of the model in RSoft is given in Ref.~\cite{Minovich:2010:PRB}. In our simulations, we use a nonuniform spatial grid with size varying from 2\,nm near the metal-dielectric interfaces to 20\,nm in free space. The time step (in the units of $ct$) is $8\times10^{-4}\,\mu$m. In order to simulate a periodic structure in the $(x,y)$ plane we use periodic boundary conditions on the planes orthogonal to $x$ and $y$ and perfectly matched layer boundary conditions on the sides orthogonal to $z$. The structure is excited from the air by a short pulse with a broad spectrum and plane wave front.
Several time monitors are used in the simulation to record the electric field components of the reflected and transmitted waves. The monitors are located $4.4\,\mu$m from the structure, so only propagating waves are recorded. We take a Fourier transform of the response and normalize it to the incident signal, compensating for the phase shift through air and the substrate. The obtained transmission $t$ and reflection $r$ which are referenced to the surfaces of the metal fishnet.

\pict[0.99]{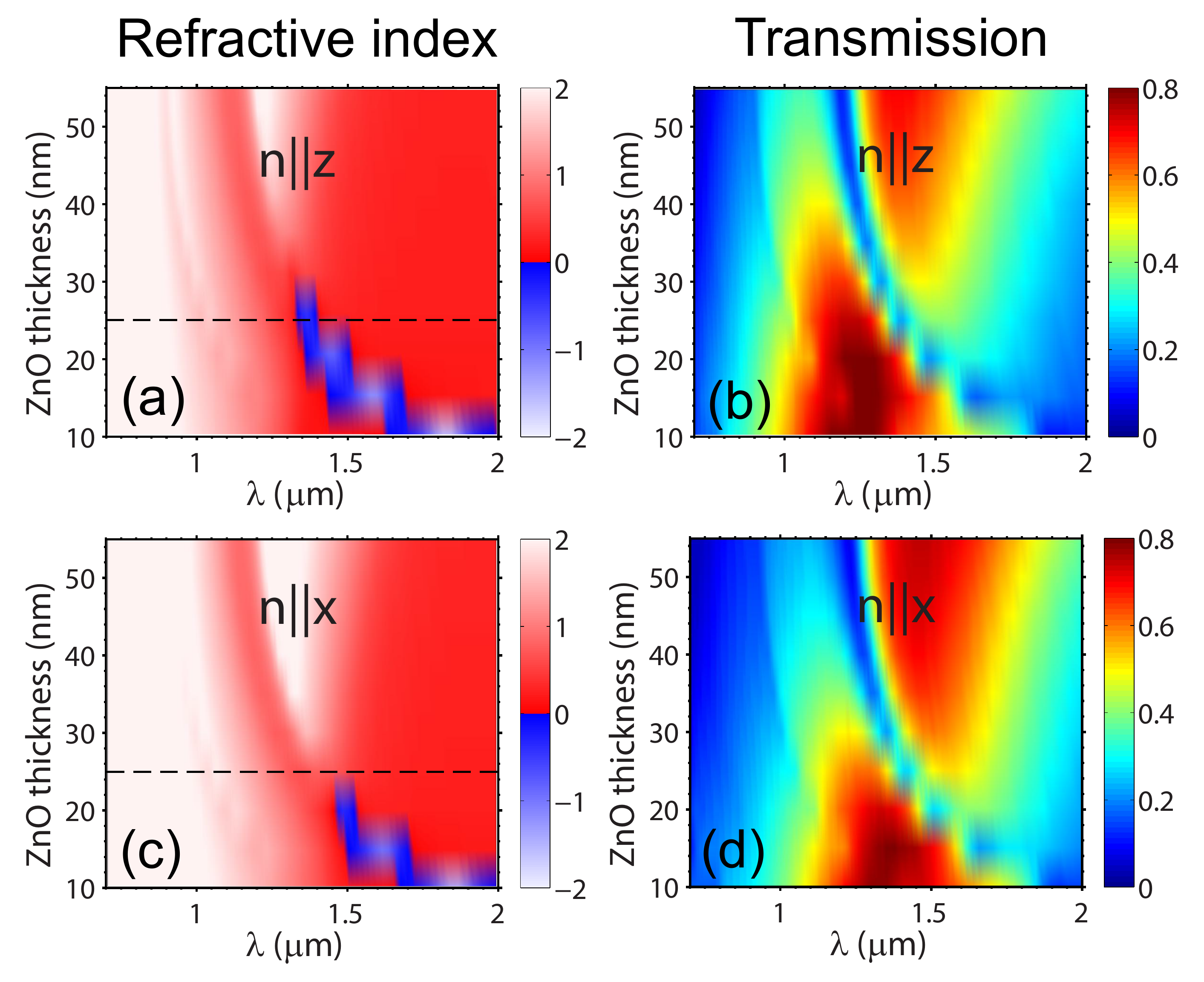}{fig2}{(color online) (a, c) Real part of the effective refractive index of the fishnet structure as a function of ZnO layer thickness. (b, d) Transmission as a function of ZnO layer thickness. Top and bottom row correspond to a director of the liquid crystal aligned along $z$ or $x$ axes, respectively.}


From the complex coefficients of transmissions and reflection, we can extract the effective refractive index $n$ and the impedance $z$ of the metamaterial structure using the Fresnel formulae~\cite{Born:1997}. For an equivalent isotropic homogeneous slab of thickness $h$ surrounded by semi-infinite media with refractive indices $n_1$ and $n_3$ and normal incidence, the expressions are
\begin{equation}
\label{z_eq}
z=\pm \left(\frac{(1+r)^2-t^2}{n_1^2(1-r)^2-n_3^2t^2}\right)^{1/2},
\end{equation}
\begin{equation}
\label{n_eq}
n=\pm\frac{1}{kh}\cos^{-1}\left\{\frac{1}{t}\frac{n_1(1-r^2)+n_3t^2}{n_1+n_3+r(n_3-n_1)}\right\}+\frac{2\pi m}{kh},
\end{equation}
where $m$ is an integer number. This result agrees with the previously derived formulae in the limit $n_1 = n_3 = 1$, and the same considerations apply to the choice of signs and the branch index $m$~\cite{Smith:2002}. We base our choice of branch on the requirement that $n$ converges in the low frequency limit, and our curves are found to agree with previous results using similar geometrical parameters~\cite{Dolling:2007, Dolling:2006}.

We then apply this extraction approach to determine the change in effective index of the metamaterial when the orientation of the director $\bm d$ is changed by $90^\circ$. We are most interested in the case when such change of the director will switch the refractive index from positive to negative. Therefore, we seek the value of the dielectric layer thickness which maximizes this effect. We perform a number of simulations varying $h_d$ and using two values of the refractive index of the liquid crystal: $n_o=1.5$ for ordinary waves when the director $\bm{d}\parallel y$ or $\bm{d}\parallel z$ (Fig.~\rpict{fig1}, left); and $n_e=1.7$ for extraordinary waves when $\bm{d}\parallel x$. These values correspond to E7 liquid crystal at $25^\circ$C~\cite{Li:2005}. We note that in the spectral range of $1-2\,\mu$m the maximum deviation of the liquid crystal refractive index from these nominal values is $0.25\%$ for $n_o$ and $0.35\%$ for $n_e$. For simplicity, we also simulate the liquid crystal as a homogeneous medium with isotropic refractive index. This is a reasonable approximation given that the electric field within the holes is mostly oriented in the same direction as the incident field~\cite{Minovich:2010:PRB}.

\pict[0.99]{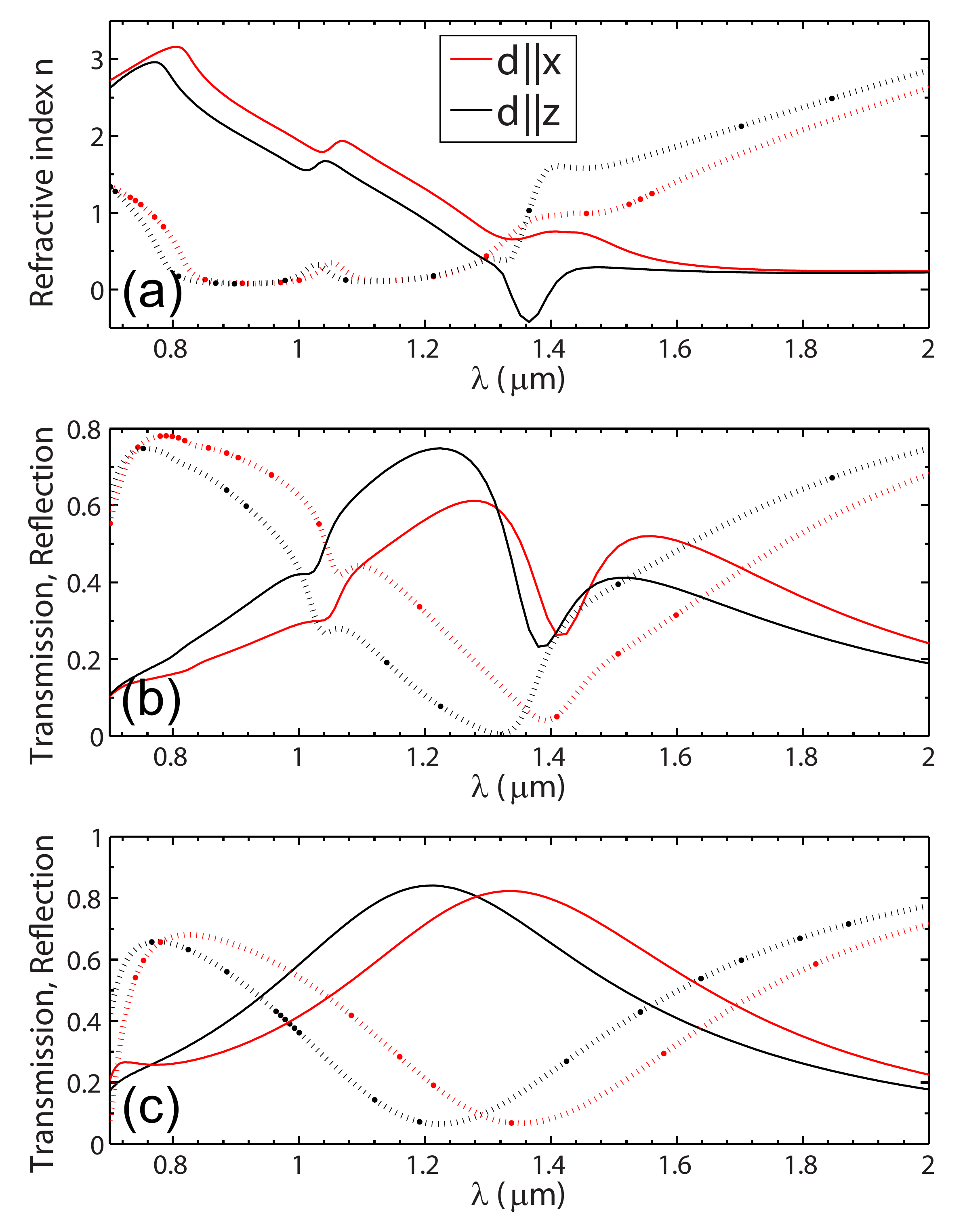}{fig3}{(color online) (a) The real (solid line) and imaginary (dotted line) parts of the effective refractive index. (b) Transmission (solid line) and reflection (dotted line). The thickness of ZnO layer $h_d=25$\,nm. (c) Transmission (solid line) and reflection (dotted line) of the respected hole array without ZnO layer.
Black lines correspond to a liquid crystal director oriented along $z$ and red lines to a director along $x$.}

The extracted effective refractive index and energy transmission coefficient $T=n_3/n_1|t|^2$ are presented in Fig.~\rpict{fig2} as a function of $h_d$ (varied in 5\,nm steps). The negative effective index in the fishnet structures is associated with the excitation of plasmons between the two metal layers~\cite{Mary:2008}, and the absorption due to these modes can be seen as transmission dips in Fig.~\rpict{fig2}(b, d). The mode excited at longer wavelengths leads to a negative refractive index for a range of thicknesses $h_d$. There is also a weakly excited mode at shorter wavelengths which can also yield to a negative index if other structural parameters are selected appropriately. Most importantly, we see that the effective refractive index switches from negative to positive value for ZnO thickness of around $25$\,nm (shown by the dashed line). This effect takes place in a narrow thickness range of only $\sim5$\,nm, however this is feasible to fabricate with contemporary sputtering techniques that can have a precision of $1$\,nm.

To understand the effects leading to switching of the effective index for $25$\,nm ZnO thickness, in Fig.~\rpict{fig3} we plot the extracted refractive index, energy transmission and reflection for both orientations of the liquid crystal director. Due to the strong spatial dispersion of the fishnet structure~\cite{Menzel:2010}, we avoid using the local effective parameters $\epsilon$ and $\mu$, since they serve only to reconstruct $n$ and $z$ and lack physical meaning of their own. Instead, we recall that the physical origin of the negative index is a gap plasmonic mode, with a strong magnetic response, lying in the region of plasma-like electrical response of the structure~\cite{Mary:2008}. In fact, this effective plasma response exists above a cut-off wavelength that is close to the cut-off wavelength of the fundamental mode of the hole array.

In order to isolate the influence of the hole modes on the response, in Fig.~\rpict{fig3}(c) we show the calculated transmission through a structure where the ZnO layer is replaced by gold, thus suppressing the gap plasmonic modes. It can be seen that the transmission peak and reflection dip in Fig.~\rpict{fig3}(b) at around $1.3\,\mu$m correspond to the cut-off resonance of the hole array, which changes drastically with the change of liquid crystal director orientation. On the other hand, the transmission dip in Fig.~\rpict{fig3}(b) that corresponds to the plasmonic dispersion curve visible in Fig.~\rpict{fig2}, barely shifts, since these modes are expected to be largely confined between the metal layers. Thus, when $n_{\rm LC}$ increases, the cut-off wavelength shifts further to the infra-red by approximately $170$\,nm and the hybridization between the hole and plasmon modes which led to the negative index is disturbed.

\pict[0.99]{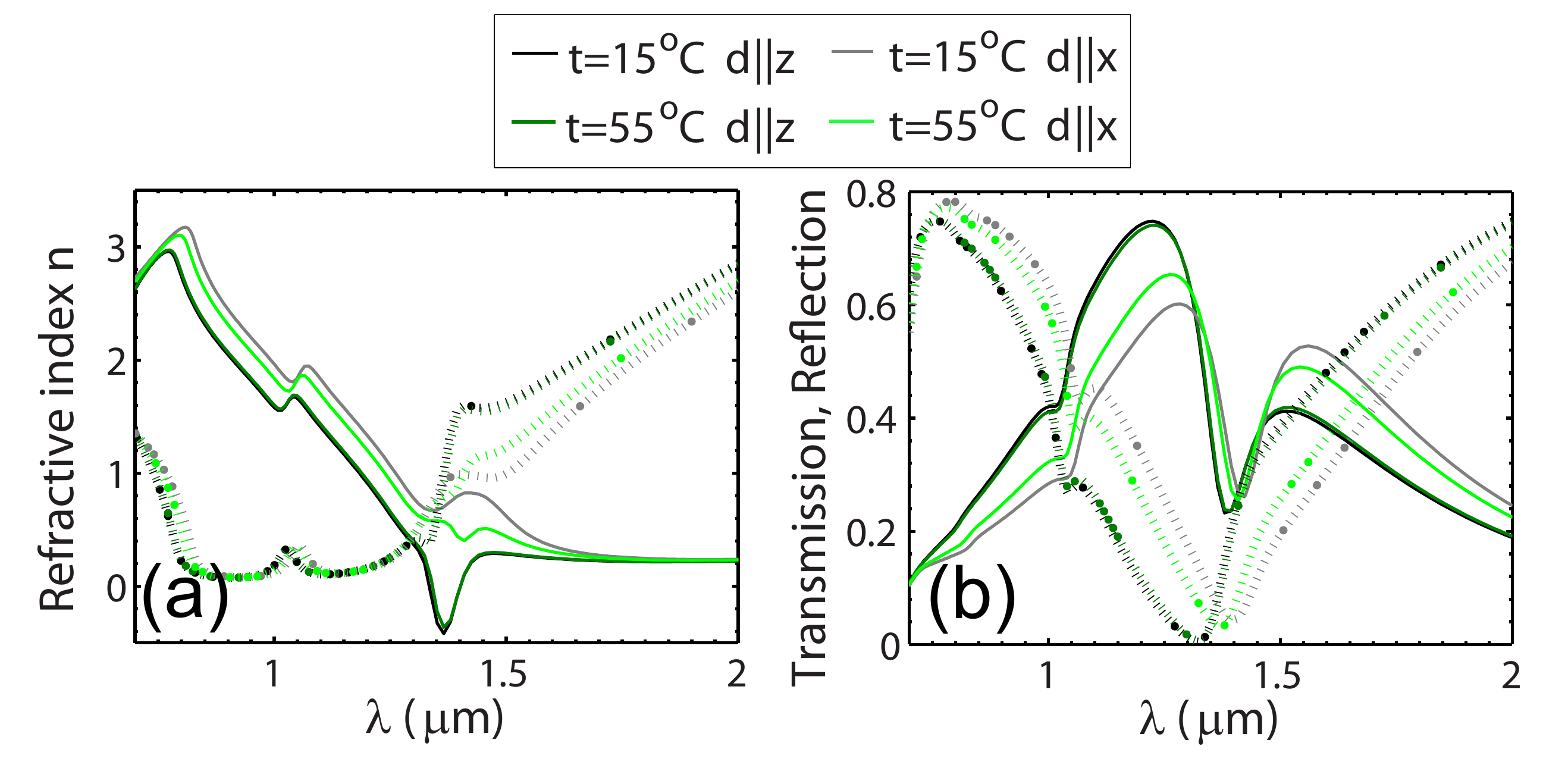}{fig4}{(color online) (a) The real (solid line) and imaginary (dotted line) parts of the effective refractive index. (b) Transmission (solid line) and reflection (dotted line).
The thickness of ZnO layer is $h_d=25$\,nm.}

We also study the effect of temperature on the infiltrated metamaterials due to the strong temperature sensitivity of liquid crystal refractive index, especially near the clearing temperature ($58^\circ$C for E7). This temperature dependence could potentially disturb the reorientation based tuning of the metamaterials. In Fig.~\rpict{fig4} we display the extracted effective refractive index, energy transmission and reflection at $15^\circ$C ($n_o=1.5$, $n_e=1.71$) and at $55^\circ$C ($n_o=1.52$, $n_e=1.65$)~\cite{Li:2005} for both orientations of the liquid crystal director. It can be clearly seen that the negative index region undergoes almost no shift with temperature, and that at both temperatures we are able to switch the index between positive and negative, thus the reorientational tuning is robust to temperature.

In conclusion, we have demonstrated numerically the tunability of optical fishnet metamaterials infiltrated by nematic liquid crystals. We have found that by tuning of the director orientation of an anisotropic liquid crystal we can reverse the sign of the effective refractive index of the fishnet metamaterials. We have shown that this tuning is dominated by a shift in the cut-off frequency of the hole modes, and that the negative index is robust to temperature variations. We believe that our results will facilitate further experimental studies of 
tunable optical metamaterials.

This work has been supported by the Australian Research Council and from the National Computing Infrastructure Merit Allocation Scheme.



\end{sloppy}

\begin{thebibliography}{99}

\bibitem{Shalaev:2007-41:NatPhot}
V.~M. Shalaev, Nat. Photonics {\bf 1}, 41 (2007).

\bibitem{Zhang:2005-137404:PRL}
S.~Zhang, W.~Fan, N.~C. Panoiu, K.~J. Malloy, R.~M. Osgood, and S.~R.~J. Brueck,
Phys. Rev. Lett. {\bf 95}, 137404 (2005).

\bibitem{Dolling:2006-231118:APL}
G.~Dolling, M.~Wegener, A.~Schaedle, S.~Burger, and S.~Linden,
Appl. Phys. Lett. {\bf 89}, 231118 (2006).

\bibitem{Chettiar:2007-1671:OL}
U.~K. Chettiar, A.~V. Kildishev, H.-K. Yuan, W.~Cai, S.~Xiao, V.~P. Drachev, and V.~M. Shalaev,
Opt. Lett. {\bf 32}, 1671 (2007).

\bibitem{Li:2007-251112:APL}
T.~Li, J.-Q. Li, F.-M. Wang, Q.-J. Wang, H.~Liu, S.-N. Zhu, and Y.-Y. Zhu,
Appl. Phys. Lett. {\bf 90}, 251112 (2007).

\bibitem{Valentine:2008-376:NAT}
J.~Valentine, S.~Zhang, T.~Zentgraf, E.~Ulin-Avila, D.~A. Genov, G.~Bartal, and X.~Zhang,
    Nature {\bf 455}, 376 (2008).

\bibitem{Minovich:2010:PRB}
A. Minovich, D.~N. Neshev, D.~A. Powell, I.~V. Shadrivov, M. Lapine, H.~T. Hattori, H.~H. Tan, C. Jagadish, and Yu.~S. Kivshar,
    Phys. Rev. B (2010) in press.

\bibitem{Lapine:2009-084105:APL}
M. Lapine, D. Powell, M. Gorkunov, I.~V. Shadrivov, R. Marques, and Yu.~S. Kivshar,
    Appl. Phys. Lett. {\bf 95}, 084105 (2009).

\bibitem{xiao_apl_09}
S. Xiao, U.~K. Chettiar, A.~V. Kildishev, V. Drachev, I.~C. Khoo, and V.~M. Shalaev,
    Appl. Phys. Lett. {\bf 95}, 033115 (2009).

\bibitem{Dicken:2009-18330:OE}
M.~J. Dicken, K. Aydin, I.~M. Pryce, L.~A. Sweatlock, E.~M. Boyd, S. Walavalkar, J. Ma, and H.~A. Atwater,
    Opt. Express {\bf 17}, 18330 (2009).

\bibitem{Zhao:2007-011112:APL}
Q. Zhao, L. Kang, B. Du, B. Li, J. Zhou, H. Tang, X. Liang, and B. Zhang,
    Appl. Phys. Lett. {\bf 90}, 011112 (2007).

\bibitem{Samson:2009:arXiv}
Z.~L. Samson, K.~F. MacDonald, F. De~Angelis, K. Knight, C.~C. Huang, E. Di~Fabrizio, D.~W. Hewak, and N.~I. Zheludev,
    arXiv:0912.4288 (2009).

\bibitem{zhang_apl_08}
F. Zhang, Q. Zhao, L. Kang, D.~P. Gaillot, X. Zhao, J. Zhou, and D. Lippens,
    Appl. Phys. Lett. {\bf 92}, 193104 (2008).

\bibitem{PowShaKiv7}
D.~A. Powell, I.~V. Shadrivov, Yu.~S. Kivshar, and M.~V. Gorkunov,
    Appl. Phys. Lett. \textbf{91}, 144107 (2007).

\bibitem{ShaKozWei8}
I.~V. Shadrivov, A.~B. Kozyrev, D.~W. van~der Weide, and Yu.~S. Kivshar,
    Appl. Phys. Lett. \textbf{93}, 161903 (2008).

\bibitem{khoo_ol_06}
I.~C. Khoo, D.~H. Werner, X. Liang, A. Diaz, and B. Weiner,
    Opt. Lett. {\bf 31}, 2592 (2006).

\bibitem{wang_apl_07}
X. Wang, D.-H. Kwon, D.H. Werner, I.C. Khoo, A.V. Kildishev, and V.M. Shalaev,
    Appl. Phys. Lett. {\bf 91}, 143122 (2007).

\bibitem{Werner:2007-3342:OE}
D.~H. Werner, D.-H. Kwon, I.~C. Khoo, A.~V. Kildishev, and V.~M. Shalaev,
    Opt. Express {\bf 15}, 3342-3347 (2007)

\bibitem{gorkunov} M. V. Gorkunov and M. A. Osipov, J. Appl. Phys. {\bf 103}, 036101 (2008).

\bibitem{Khoshman:2007}
J.~M. Khoshman and M.~E. Kordesch,
    Thin Solid Films {\bf 515}, 7393 (2007).

\bibitem{Rakic:1998-5271:AO}
A.~D. Rakic, A.~B. Djurisic, J.~M. Elazar, and M.~L. Majewski,
Appl. Opt. {\bf 37}, 5271 (1998).

\bibitem{Born:1997}
M. Born, E. Wolf, A.~B. Bhatia, {\it Principles of Optics}, 61-70 (Cambridge University Press, Cambridge, 1997).

\bibitem{Smith:2002}
D.~R. Smith, S. Schultz, P. Markos, P. and C.~M. Soukoulis,
  Phys. Rev. B {\bf 65}, 195104 (2002).

\bibitem{Dolling:2007}
G. Dolling, M. Wegener, C.~M. Soukoulis, and S. Linden,
    Opt. Exp. {\bf 15}, 11536 (2007).

\bibitem{Dolling:2006}
G. Dolling, C. Enkrich, M. Wegener, C.~M. Soukoulis, and S. Linden,
    Opt. Lett. {\bf 31}, 1800 (2006).

\bibitem{Li:2005}
J. Li, S.-T. Wua, S. Brugioni, R. Meucci, and S. Faetti,
    J. Appl. Phys. {\bf 97}, 073501 (2005).

\bibitem{Mary:2008}
A. Mary, S.~G. Rodrigo, F.~J. Garcia-Vidal and L. Martin-Moreno,
Phys. Rev. Lett. {\bf 101}, 103902 (2008).

\bibitem{Menzel:2010}
C. Menzel, T. Paul, C. Rockstuhl, T. Pertsch, S. Tretyakov and F. Lederer,
Phys. Rev. B {\bf 81}, 035320 (2010).






\end{thebibliography}
\end{document}